\documentclass[twocolumn,showpacs,preprintnumbers,amsmath,amssymb,floatfix,prd]{revtex4}
\usepackage{epsfig}
\usepackage{dcolumn}
\begin{document}
\title{Unveiling the Proton Spin Decomposition at a Future Electron-Ion Collider}
\author{Elke C.\ Aschenauer}\email{elke@bnl.gov}
\affiliation{Physics Department, Brookhaven National Laboratory, Upton, NY~11973, USA}
\author{Rodolfo Sassot}\email{sassot@df.uba.ar}
\affiliation{ Departamento de F\'{\i}sica and IFIBA,  Facultad de Ciencias Exactas y Naturales, Universidad de Buenos Aires, Ciudad Universitaria, Pabell\'on\ 1, (1428) Buenos Aires, Argentina}
\author{Marco Stratmann}
\email{marco.stratmann@uni-tuebingen.de}
\affiliation{Institute for Theoretical Physics, University of T\"ubingen, Auf der Morgenstelle 
14, 72076 T\"ubingen, Germany}

\begin{abstract}
We present a detailed assessment of how well a future Electron-Ion Collider 
could constrain helicity parton distributions in the nucleon and, therefore, 
unveil the role of the intrinsic spin of quarks and gluons in the proton's spin budget.
Any remaining deficit in this decomposition will provide the best indirect constraint on the 
contribution due to the total orbital angular momenta of quarks and gluons.
Specifically, all our studies are performed in the context of global QCD analyses
based on realistic pseudo-data and in the light of the most recent data obtained from polarized proton-proton collisions at 
BNL-RHIC that have provided evidence for a significant gluon polarization
in the accessible, albeit limited range of momentum fractions.
We also present projections on what can be achieved on the gluon's helicity distribution
by the end of BNL-RHIC operations. All estimates of current and projected uncertainties
are performed with the robust Lagrange multiplier technique.
\end{abstract}

\pacs{13.88.+e, 13.60.Hb, 12.38.Bx}

\maketitle

\section{Introduction and Motivation}
%
The exploration of the nucleon's inner structure and the interactions 
among its constituents at high energies has been the main protagonist in the 
quest for a quantitative picture of matter at the most elementary level 
during the last fifty years. 
In spite of an impressive wealth of achievements and discoveries, 
many fundamental questions still remain unanswered today. 
The origin of the proton spin is a remarkable example for one of these
compelling questions still driving the field of Nuclear Physics.
Inspired by the Quark Parton Model, it was originally thought to arise solely from the
intrinsic spins of the proton's valence up and down quarks,
but this naive view was denied by results from polarized deep inelastic scattering (DIS) 
experiments in the late eighties \cite{Aidala:2012mv}
and remained under experimental and
theoretical scrutiny and debate ever since.

The proposed Electron-Ion Collider (EIC) project in the U.S.\ 
\cite{Boer:2011fh,Accardi:2012qut,Aschenauer:2014cki}, 
a versatile machine designed to explore nuclei and polarized light ions at the 
energy and intensity frontier
with the precision of electromagnetic and electroweak probes, will be the next milestone in the quest for 
gaining a deeper insight into the nucleon's inner workings.
Among its remarkable and unprecedented capabilities, an EIC will be able to disclose 
definitively the role of quark and gluon spins in the proton down to fractions $x$ of 
a few hundred thousandths of the nucleon's momentum 
and in a wide range of resolution scales set by virtuality $Q$ of the probing photon
in inclusive deep-inelastic scattering (DIS).
Once the intrinsic spin contributions are being tightly constrained, an EIC will deliver early on
a first quantitative, though indirect, estimate on the combined orbital angular momenta (OAM) of 
quarks and gluons needed in the decomposition of the proton spin.
Of course, an EIC is also aiming at directly accessing OAM \cite{Boer:2011fh,Accardi:2012qut,Aschenauer:2014cki,Aschenauer:2013hhw} which is, however, experimentally much more challenging
than inclusive DIS and, hence, will require some years of running and additional
theoretical groundwork to bear fruit.

In the phenomenologically very successful framework of perturbative QCD (pQCD) 
information on how momentum and spin of quarks and gluons are apportioned in the nucleon 
factorizes in a universal, process-independent way from calculable, short-distance
partonic scattering cross sections \cite{Collins:1989gx}.
In case the incident beams are both longitudinally polarized in the scattering process, 
helicity parton distributions (PDFs) $\Delta f(x,Q^2)$ are accessible, 
where $f$ denotes the different (anti-)quark flavors $q$ $(\bar{q})$ or the gluon $g$.
For factorization to be a viable approach, the external scale $Q$ characterizing the scattering needs to
be large enough, say, above about $1-2\,\mathrm{GeV}$.
Upon integration over all momentum fractions $x$, 
\begin{equation}
\label{eq:firstmom}
\Delta f(Q^2) = \int_0^1 dx \Delta f(x,Q^2)
\end{equation}
the helicity PDFs
contain the desired information entering in the decomposition of the proton's spin
\begin{equation}
\label{eq:spinsum}
\frac{1}{2} = \frac{1}{2} \sum_{q} \left[\Delta q(Q^2) + \Delta\bar{q}(Q^2)\right]
+ \Delta g(Q^2) + {\cal{L}}(Q^2)
\end{equation}
where ${\cal{L}}(Q^2)=\sum_q [L_q(Q^2)+L_{\bar{q}}(Q^2)]+L_g(Q^2)$ is the 
total contribution from OAM.
The sum of quark and antiquark densities in Eq.~(\ref{eq:spinsum}) can be combined in the quark
singlet contribution $\Delta\Sigma(Q^2)$.
It should be noted that the apparently simple relation (\ref{eq:spinsum})
was subject to quite some debate and controversy in recent years. 
The main problems are that the decomposition (\ref{eq:spinsum}) is not unique
and that the separation of the gluonic total angular momentum into $\Delta g(Q^2)$
and $L_g(Q^2)$ was not thought to be possible in a gauge invariant manner.
It is now understood \cite{Leader:2013jra} that a particular choice of decomposition
is essentially a matter of taste as long as each component can be determined, at least in principle, 
experimentally or from lattice QCD calculations. Each variant covers complementary aspects
of a complex bound-state system such as the nucleon.

In this paper, we are primarily interested in providing the best possible quantitative assessment 
of the impact a future EIC would have on determinations of the 
quark singlet and gluon helicity densities and their contributions to the proton spin
in Eq.~(\ref{eq:spinsum}).
The theoretical framework for our survey is based on an earlier study \cite{Aschenauer:2012ve}
where we have performed a series of global QCD analyses of helicity PDFs at 
next-to-leading order (NLO) accuracy including sets of realistic mock EIC data for both polarized DIS and semi-inclusive DIS (SIDIS).
These data were generated and properly randomized within one sigma uncertainties
using state-of-art spin-dependent PDFs \cite{deFlorian:2008mr,deFlorian:2009vb}
based on experimental information from polarized fixed target DIS and SIDIS
and proton-proton collisions at BNL-RHIC available at that time.

Since then, many new and important results by the BNL-RHIC experiments 
\cite{Adare:2008qb,Adare:2008aa,Adamczyk:2012qj,Adamczyk:2013yvv,Adamczyk:2014ozi,Adare:2015oqa}  
have been reported, that have changed substantially our perception of the 
gluon helicity distribution \cite{deFlorian:2014yva}.
QCD analyses of first RHIC spin results \cite{deFlorian:2008mr,deFlorian:2009vb}
were dominated by single-inclusive pion production data, 
obtained at rather low transverse momentum scales $p_T$, and
suggested very little or no gluon polarization in the 
range $0.05\lesssim x \lesssim 0.2$ predominantly probed by data,
albeit within very large uncertainties.
In contrast, the most recent and very precise jet production data \cite{Adamczyk:2014ozi}
from RHIC, taken at larger scales $p_T$ than for pions, provided for the first time
clear evidence for a sizable gluon polarization \cite{deFlorian:2014yva};
very similar results have been reported in an independent global QCD analysis
in Ref.~\cite{Nocera:2014gqa}.
It is important to notice that both sets of recent inclusive measurements at RHIC
which are highly sensitive to the gluon helicity PDF, jets \cite{Adamczyk:2014ozi}
and pions \cite{Adare:2015oqa}, are fully consistent with each other
if one properly takes into account the different scales $p_T$ and 
ranges of momentum fractions $x$ probed by the respective data. This is automatically
guaranteed and achieved in the framework of a global QCD analysis at NLO accuracy
based on exact, unabridged NLO expressions for the underlying spin-dependent $pp$ cross sections
\cite{Jager:2002xm,Jager:2004jh}. 
   
In light of these new results from RHIC pertaining to the relevance of gluons in the nucleon's
spin decomposition it is of critical importance to reexamine our previous 
impact study \cite{Aschenauer:2012ve} to update the case for a future polarized DIS 
program at an EIC.
Apart from including the latest RHIC spin data we will also refine and expand our study in many
other important aspects. 
To establish a meaningful baseline for estimating the impact of future EIC DIS data, 
we first present projections of what will likely be the
final word from the RHIC spin program with respect to $\Delta g(x,Q^2)$ and its integral
$\Delta g(Q^2)$. To this end, we add realistic projections for yet to be published 
inclusive jet and pion data from polarized $pp$ collisions at $200$ and $510\,\mathrm{GeV}$ 
\cite{Aschenauer:2015eha} to the global QCD analysis framework of Ref.~\cite{deFlorian:2014yva}.
First, preliminary results \cite{ref:rhic-upcoming} for these measurements
are fully consistent with the most recent global fits of
helicity PDFs \cite{deFlorian:2014yva,Nocera:2014gqa}.
Next, we update and include the sets of mock polarized DIS data to reflect the latest 
collision energies envisioned for the eRHIC option of an EIC \cite{Aschenauer:2014cki}.
Throughout, we pay special attention to properly estimate the impact of 
the different data sets in the various regions of parton momentum fraction $x$.

Once we have completed our extensive studies of the prospects for the 
gluon helicity density $\Delta g(x,Q^2)$ and its integral $\Delta g(Q^2)$
from both the RHIC spin program and a future EIC, we will consider
the quark spin contribution $\Delta \Sigma(x,Q^2)$ summed over all flavors. 
Finally, we will examine what can be gleaned from these two projections
for the total parton OAM contribution ${\cal{L}}(Q^2)$ when combined with
the spin decomposition (\ref{eq:spinsum}).

As a technical improvement, we introduce in all our impact studies 
a more robust treatment of uncertainties.
Specifically, we now estimate uncertainties with the Lagrange multiplier technique \cite{Stump:2001gu} 
considering dynamical tolerances corresponding to 
the $90$\% confidence level (C.L.) \cite{deFlorian:2014yva}
as it is customary in most current fits of unpolarized parton densities \cite{ref:pdf-recent}, 
rather than adopting some ad hoc fixed $\Delta \chi^2$ criterion 
as in our previous analyses \cite{deFlorian:2008mr,deFlorian:2009vb,Aschenauer:2012ve}.
This new procedure takes into account more precisely the role of the different 
observables in constraining parton densities at different kinematic regions in
$x$ and $Q^2$.
 
We have to postpone a new detailed study of the flavor separated helicity quark
distributions $\Delta f(x,Q^2)$ as a proper global QCD analysis of 
projected polarized SIDIS data with identified pions and kaons
for an EIC critically depends on having available a reliable set of
parton-to-hadron fragmentation functions (FFs) and their uncertainties.
At this point, only pion FFs have been updated
recently \cite{deFlorian:2014xna} and kaon FFs,
important for determining the strangeness polarization,
are still work in progress \cite{ref:kaonffs}.  
Also, RHIC spin data, this time for the longitudinal 
single spin asymmetry of $W^{\pm}$ bosons \cite{Aschenauer:2015eha}
are expected to play a significant role in establishing a meaningful baseline for 
impact studies of flavor separated $\Delta f(x,Q^2)$ from SIDIS measurements 
at an EIC.
However, the final data sets from RHIC are not yet available.
Finally, we note that for this study we focus in general on the 
so far unmeasured region of small momentum fractions 
and, correspondingly, low-to-medium values of $Q^2$ at an EIC. The prospects of DIS measurements
with an EIC in the region where $Q\simeq M_{W}$ was studied in some 
detail in Ref.~\cite{Aschenauer:2013iia} based on NLO expressions 
for charged current DIS given, for instance, in \cite{deFlorian:1994wp}.

In the remainder of the paper, we first briefly discuss the sets of pseudo-data generated to
study the impact of both polarized DIS at an EIC and upcoming RHIC $pp$ data
in determinations of the gluon and quark singlet helicity distribution and their 
respective integrals. In Secs.~III and IV we present the results of our global QCD analyses 
at NLO accuracy based on including these sets of data one-by-one. First, we focus on
what can be expected for $\Delta g(x,Q^2)$ by the end of the
RHIC spin program and, next, we include also the mock EIC data.
Our main results are best presented in terms of ``running $x$-integrals'' for 
$\Delta g$ (in Sec.~III), $\Delta \Sigma$, and the difference between $1/2$ and  
$\frac{1}{2} \Delta\Sigma+\Delta g$ (in Sec.~IV), 
that represents roughly how much room is left for OAM to saturate the spin sum rule.  
Finally, we summarize our results and present our conclusions.

\section{Simulated Data for polarized DIS at an EIC}
%
Inclusive DIS with longitudinally polarized leptons and nucleons can be
expressed by the structure function $g_1(x,Q^2)$. For energy scales $Q$ well below the
mass of the electroweak gauge bosons $W^{\pm}$ and $Z$, the scattering is mediated by the
exchange of a virtual photon and both kinematic variables $x$ and $Q^2$ can
be straightforwardly reconstructed from measuring the energy and angle of the deflected lepton.

Sensitivity to the gluon helicity distribution $\Delta g(x,Q^2)$ in DIS
is manifold: directly, through higher order QCD corrections to $g_1$, which are
sub-leading compared to the dominant quark contribution and, hence, small
and difficult to utilize in an analysis, or, indirectly, through scaling violations, that is variations of the structure function $g_1(x,Q^2)$ with scale $Q$ for {\em fixed} values of $x$.
The smaller $x$ the more pronounced are the scaling violations, which are
usually expressed as the logarithmic derivative of the DIS structure function, i.e.,
$dg_1(x,Q^2)/d\,\ln(Q^2)$.
It should be noted that
corresponding results for unpolarized DIS from the DESY-HERA experiments \cite{Abramowicz:2015mha}
are utilized in all current extractions of helicity-averaged PDFs and provide by far the
best constraint on the gluon density below about $x\simeq 0.05$ \cite{ref:pdf-recent}.
Through evolution, $\Delta g(x,Q^2)$ is also correlated with the quark singlet helicity distribution,
at small $x$ mainly to the sea component, which itself is very poorly constrained 
below the range of existing polarized DIS data covering $x\gtrsim 0.004$ for $Q^2>1\,\mathrm{GeV}^2$
\cite{Aidala:2012mv,Adolph:2015saz}.
As we shall see below, this also leads to a significant uncertainty in current estimates of
$g_1(x,Q^2)$ and, hence, in the quark singlet $\Delta \Sigma(Q^2)$ entering (\ref{eq:spinsum}),
which only an EIC can finally resolve. 
At variance with unpolarized PDFs, where the scale evolution at small $x$ is driven by gluons, 
quark and gluon helicity PDFs  are equally important as they exhibit a similar, less singular $x\to 0$ behavior  in all evolution kernels \cite{ref:pol-evol}. The different evolution of helicity and helicity-averaged 
distributions also leads to a strong suppression of gluon and quark polarizations as $x\to 0$ and,
consequently, of all spin asymmetries in that kinematic region. 

Thus, in order to get an accurate picture of the gluon and quark helicity densities and,
in particular, their $x$-integrals (\ref{eq:firstmom}) at an EIC it is not only crucial 
to have good precision in a wide range of parton momentum fractions $x$ 
to reduce extrapolation uncertainties from the currently unmeasured
$x$ region in DIS, but also to cover for any given $x$ the largest possible range 
in the photon virtuality $Q^2$ to determine $\Delta g$ from scaling violations.
The requirement of being able to reach $x$ values
well below of what has been achieved so far \cite{Aidala:2012mv,Adolph:2015saz} for virtualities
$Q^2$ safely in the DIS regime, say, above at least $1\,\mathrm{GeV}^2$, clearly
calls for the largest conceivable center-of-mass system (c.m.s.) energies $\sqrt{s}$ 
at a future $ep$ collider in order to make any significant impact. 
Furthermore, at small $x$ the reach in $Q^2$ to study scaling violations, 
which are only logarithmic in energy, is kinematically
limited to $Q^2=s\,x\,y$ where $y$ is the fractional energy of the virtual photon.
$y$ is constrained by the increasing depolarization of the virtual photon from below, 
which should not exceed about $90\%$, and, from above, by the energy of the scattered lepton, 
which should be not too low to be still reliably measurable in the detector. 
In practice, $0.01\lesssim y \lesssim 0.95$
appears to be a reasonable choice as the DESY HERA collider experiments even
used the range $0.005\lesssim y \lesssim 0.95$ for parts of their physics program.

%
\begin{table}[bth!]
\caption{\label{tab:kine} Combinations of electron and proton energies of the current
eRHIC design \cite{Aschenauer:2014cki} as used in our analyses, the corresponding c.m.s.\ energies, 
and the range in $x$ accessible for $Q^2=1\,\mathrm{GeV}^2$ and assuming $ 0.01 < y < 0.95$.}
\begin{ruledtabular}
\begin{tabular}{ccccc}
$E_e\times E_p$   & $\sqrt{s}$       & $x_{\min}$          & $x_{\max}$\\
$[\mathrm{GeV}]$  & $[\mathrm{GeV}]$ & for $y=0.95$ & for $y=0.01$ \\ \hline
$15\times 100$    & 77.5             & $1.8\times 10^{-4}$ &  $1.7\times 10^{-2}$ \\
$15\times 250$    & 122.7            & $7.0\times 10^{-5}$ &  $6.7\times 10^{-3}$ \\
$20 \times 250$   & 141.4            & $5.3\times 10^{-5}$ &  $5.0\times 10^{-3}$\\
\end{tabular}
\end{ruledtabular}
\end{table}
In our previous study \cite{Aschenauer:2012ve}, we adopted c.m.s.\ energies from $\sqrt{s}\simeq 45\,\mathrm{GeV}$ to $71\,\mathrm{GeV}$ corresponding to an electron beam energy of $5\,\mathrm{GeV}$ 
for the BNL eRHIC design of an initial version of an EIC at that time.  
As a potential later upgrade we also considered $E_e=20\,\mathrm{GeV}$ leading to
a maximum $\sqrt{s}=141.4\,\mathrm{GeV}$.
The latest eRHIC design evolved considerably and now
has a default electron beam energy of about $15\,\mathrm{GeV}$ 
but is also capable to run with both lower and higher lepton energies at similar
luminosities \cite{Aschenauer:2014cki} from day one.
When combined with the existing RHIC proton beam of up to $E_p=250\,\mathrm{GeV}$ 
this set-up not only extends the reach towards small $x$ 
by almost two decades compared to what is known today
but also makes the detection of the scattered electron at large $y$ much easier
than for a $5\,\mathrm{GeV}$ electron beam.
Table~\ref{tab:kine} summarizes the three different c.m.s.\ energies we are going to
consider in this study along with the accessible range in $x$, $[x_{\min},x_{\max}]$,
for $Q^2=1\,\mathrm{GeV}^2$,
assuming the same standard DIS cuts as in our previous analysis \cite{Aschenauer:2012ve}:
$Q^2>1\,\mathrm{GeV}^2$, $W^2>10\,\mathrm{GeV}^2$, and $0.01\le y \le 0.95$.
As can be seen, the improvements in the kinematic range for both $x$ and $Q^2$ from $\sqrt{s}=122.7$ to
$141.4\,\mathrm{GeV}$ corresponding to $E_e=15\,\mathrm{GeV}$ 
and $20\,\mathrm{GeV}$, respectively, are not too significant.

In order to generate the mock polarized DIS data for the energies shown in
Tab.~\ref{tab:kine}, we proceed very much in the same way as was 
detailed in Ref.~\cite{Aschenauer:2012ve}.
The statistical accuracy for each of the generated data sets is scaled to correspond to a very modest
accumulated integrated luminosity of $10\,\mathrm{fb}^{-1}$, equivalent to at most a few months
of operations for the anticipated luminosities for eRHIC \cite{Aschenauer:2014cki} and
$50\%$ efficiency in the data taking.
We assume $70\%$ polarization for both the electron and the proton beam.
Each of the three new sets adds about 70 points to the existing suite 
of fixed target DIS spin asymmetry data, 
distributed logarithmically in 4 [5] bins per decade in $Q^2$ $[x]$. As in
Ref.~\cite{Aschenauer:2012ve}, the actual data used 
in our global analyses below are not the generated ones but 
theoretical estimates of the spin asymmetry
at NLO accuracy based on the latest DSSV helicity densities \cite{deFlorian:2014yva}, 
reflecting the same relative statistical
accuracy in each $x, Q^2$ bin as the Monte Carlo data, and having their central values 
randomized within one-sigma uncertainties.

In any case, the DIS measurements at an EIC will be mainly limited by systematical errors
due to detector performance, beam polarization and luminosity measurements, and the unfolding
of QED radiative corrections, which we all assume to be controlled to a percent level accuracy.
This will be necessary for achieving meaningful constraints on the helicity PDFs as we have 
already demonstrated in our previous analysis, see, e.g., Fig.~6 in Ref.~\cite{Aschenauer:2012ve}. 
We note that the typical size of the DIS spin asymmetries at the lowest values of $x$ 
can be as small as a few times $10^{-4}$.
Such measurements are already routinely performed at RHIC in case of double spin asymmetries
for single-inclusive pion production \cite{Adare:2015oqa} at transverse momenta $p_T$ of a 
few $\mathrm{GeV}$.

%
\begin{figure}[thb!]
\vspace*{-0.4cm}
\begin{center}
\vspace*{-0.6cm}
\epsfig{figure=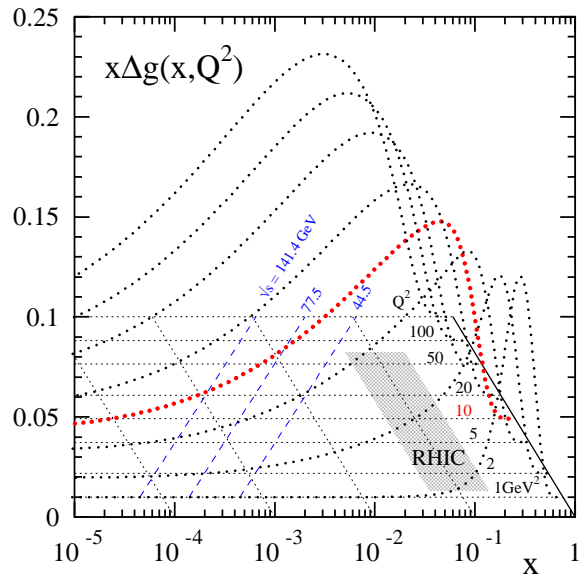,width=0.52\textwidth}
\end{center}
\vspace*{-0.7cm}
\caption{\label{fig:deltag} [color online]
QCD scale evolution of the optimum DSSV 2014 gluon helicity distribution \cite{deFlorian:2014yva}
as a function of the momentum fraction $x$ for various values of $Q^2$ (dotted curves).
The dashed lines delineate the minimal $x$ accessible at an EIC in the $x-Q^2$ plane
for various c.m.s.\ energies and $y=0.95$.
The line for $\sqrt{s}=44.5\,\mathrm{GeV}$ corresponds to an energy considered in our previous paper
\cite{Aschenauer:2012ve} utilizing only a $5\,\mathrm{GeV}$ electron beam at eRHIC.
The shaded area labeled ``RHIC'' shows approximately the region where $\Delta g(x,Q^2)$
is constrained best by published RHIC $pp$ data \cite{Adamczyk:2014ozi,Adare:2015oqa}.}
\end{figure}
To illustrate the main idea of the planned measurements in polarized DIS,
the enhanced kinematic coverage at an EIC, and the current uncertainties 
which prevent us from having a clear picture of 
how quarks and gluons generate the proton's spin, 
we start with presenting the scale dependence of the gluon helicity density.
Figure~\ref{fig:deltag} shows the shape of $\Delta g(x,Q^2)$ as 
extracted from the latest global QCD analysis to all presently available data in DIS, SIDIS, and
$pp$ collisions \cite{deFlorian:2014yva} (henceforth denoted as ``DSSV 2014 best fit''). 
The plot illustrates how the QCD scale evolution
quickly broadens the $x$-shape of $\Delta g (x,Q^2)$ 
with increasing $Q^2$, which is initially largely concentrated at rather high 
momentum fractions $x$ at a scale of $Q^2=1\,\mathrm{GeV}^2$, and moves
its peak rapidly towards smaller values of $x$.
The main constraint on $\Delta g(x,Q^2)$ so far is derived from the 
latest RHIC inclusive jet and $\pi^0$
\cite{Adamczyk:2014ozi,Adare:2015oqa} which span, as is indicated in Fig.~\ref{fig:deltag}, 
only a rather limited range in both $x$ and $Q^2$, the latter being approximately determined by 
the measured range of $p_T$ (a few GeV to about 20-30 GeV). 
Clearly, nothing is known about the low momentum fraction tail of $\Delta g(x,Q^2)$ 
beyond the part of it generated radiatively by evolution from lower $Q^2$ and larger $x$. 
In the next Section, we will elaborate in detail on the present uncertainties for $\Delta g(x,Q^2)$,
which appear to very substantial, in particular, in the unmeasured small $x$ region.
In general, spin asymmetries sensitive to small $x$ are
very difficult to seize experimentally as the unpolarized gluon density rises sharply 
as $x\to 0$, leading to a more and more suppressed gluon polarization $\Delta g/g$, likewise
for sea quark polarizations.

\begin{figure}[tbh!]
\begin{center}
\vspace*{-0.6cm}
\epsfig{figure=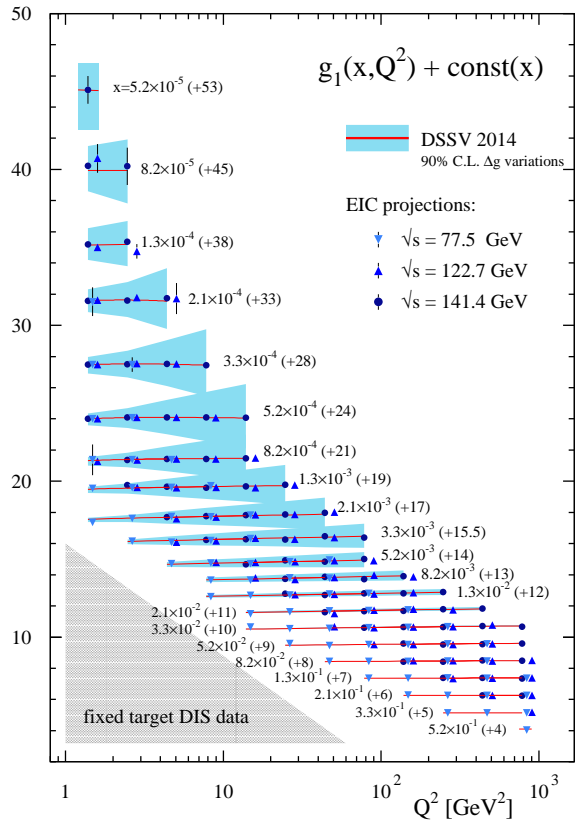,width=0.5\textwidth}
\end{center}
\vspace*{-0.5cm}
\caption{\label{fig:g1} [color online]
Projected EIC data for the structure function $g_1(x,Q^2)$ for the
different combinations of electron and proton energies in Tab.~\ref{tab:kine}.
Constants are added to $g_1$ to separate the different $x$ bins and multiple data points 
in the same $(x,Q^2)$ bin are slightly displaced horizontally.
The solid lines are obtained for the optimum DSSV fit of 2014 \cite{deFlorian:2014yva}
and the shaded bands illustrate the 90$\%$ C.L.\ uncertainties due to variations
in the {\em gluon helicity density}.
The shaded region in the lower left corner illustrates the $(x,Q^2)$ region covered
by present fixed target data.}
\end{figure}
To emphasize the importance of running an EIC at the largest possible c.m.s.\ energy, 
we also overlay in Fig.~\ref{fig:deltag}, the kinematic lower limits in $x$ and $Q^2$ 
for various values of $\sqrt{s}$ and assuming $y\le 0.95$, see Tab.~\ref{tab:kine}.
The line for $\sqrt{s}=44.5\,\mathrm{GeV}$ corresponds to the no longer considered lowest c.m.s.~energy
of our previous study \cite{Aschenauer:2012ve}; as was mentioned above, the new design for eRHIC 
allows to cover significantly lower values of $x$ from the very beginning of operations.

%
%
\begin{figure}[thb!]
\begin{center}
\vspace*{-0.9cm}
\epsfig{figure=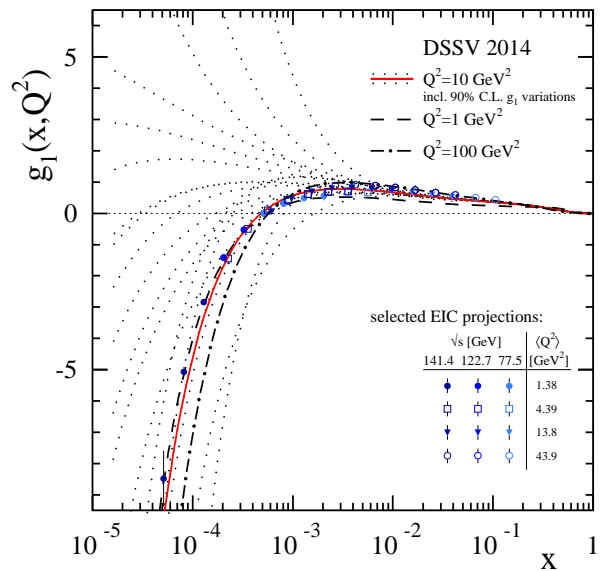,width=0.52\textwidth}
\end{center}
\vspace*{-0.7cm}
\caption{\label{fig:g1cloud} [color online] The polarized DIS structure function $g_1(x,Q^2)$ 
at $Q^2=10\,\mathrm{GeV}^2$
as a function of $x$ computed with the optimum DSSV 2014 helicity PDFs \cite{deFlorian:2014yva}
(solid line). The dotted curves represent alternative fits within 90\% C.L.\ 
uncertainties.
The dashed and dot-dashed lines show the effects of the scale evolution from
$Q^2=1\,\mathrm{GeV}^2$ to $100\,\mathrm{GeV}^2$.
The points illustrate typical uncertainties and the kinematic reach of projected EIC data
for the three different c.m.s.\ energies listed in Tab.~\ref{tab:kine}.}
\end{figure}
Figure~\ref{fig:g1} illustrates our updated simulated data sets for inclusive polarized DIS at an EIC
for the three different choices of c.m.s.\ energies listed in Tab.~\ref{tab:kine}. 
The solid lines reflect the expectations from the best fit of DSSV 2014 \cite{deFlorian:2014yva}
by extrapolating their results outside the experimentally constrained $x$ and $Q^2$ range.
The shaded bands illustrate the uncertainty estimates corresponding to the $90$\%~C.L.~variations 
of $\Delta g(x,Q^2)$ given in Ref.~\cite{deFlorian:2014yva},
which cover a very significant spread below about $x\simeq 0.01$; see also Fig.~1 in 
Ref.~\cite{deFlorian:2014yva}. 
The error bars for the EIC pseudo-data were determined as described above 
and in Ref.~\cite{Aschenauer:2012ve}
and reflect the expected statistical accuracy for a modest integrated luminosity of $10\,\mathrm{fb}^{-1}$,
$70\%$ beam polarization, and $50\%$ efficiency in the data taking.
We recall that all currently available polarized DIS data cover only the lower left corner
in Fig.~\ref{fig:g1} with the smallest $x$, $x\simeq 3.6\times 10^{-3}$, being reached by the
recent COMPASS data \cite{Adolph:2015saz} for $Q^2\simeq 1\,\mathrm{GeV}^2$.
As can be seen, in the kinematic region already covered well by present fixed target data,
$x\gtrsim 0.01$, the remaining uncertainties in $g_1(x,Q^2)$ are very small. 
For smaller $x$, the precision of the projected EIC data is significantly better than
current uncertainties and these measurements will be the decisive factor in future global fits
as we shall illustrate in the next Section.

One notices the rather modest scaling violations $dg_1(x,Q^2)/d\ln Q^2$ 
for the optimum DSSV 2014 fit throughout the entire
$x$ and $Q^2$ range shown in Fig.~\ref{fig:g1}, in particular, if compared to similar plots
for the unpolarized DIS structure functions \cite{Abramowicz:2015mha}.
On the one hand, this is due to the less singular scale evolution for helicity PDFs at small $x$,
and, on the other hand, there is also a potential delicate cancellation with the
quark helicity PDFs, which, as $\Delta g$ itself, are not bound to be positive definite and, 
in addition, can have different
signs for different flavors. Therefore, alternative fits, like those for $\Delta g$ shown in Fig.~\ref{fig:deltag}, will all exhibit somewhat different patterns of
scaling violations than the optimum DSSV 2014 fit.

As we shall see next, our current ignorance of the small $x$ behavior of helicity quark densities
also imposes a significant uncertainty on expectations for $g_1(x,Q^2)$ in the EIC regime.
In Fig.~\ref{fig:g1cloud} we present the DIS structure function
$g_1(x,Q^2)$ (solid line) and $90\%$ C.L.\ estimates of its uncertainties (dotted lines)
as a function of the momentum fraction $x$ at $Q^2=10\,\mathrm{GeV}^2$.
Unlike in Fig.~\ref{fig:g1}, the alternative fits at $90\%$ C.L. now include
{\em combined} variations of quark and gluon helicity PDFs away from the 
DSSV 2014 best fit \cite{deFlorian:2014yva}
which lead to uncertainties at least twice as large as for the variations just based on
$\Delta g$ shown in Fig.~\ref{fig:g1}. We note that throughout our paper
the allowed ranges of variations at $90\%$ C.L.\ are determined for each of the shown results
by the robust Lagrange multiplier technique and dynamic tolerances for the 
appropriate increase in the $\chi^2$ of the fit similar to what is done in 
most of the recent PDF fits \cite{ref:pdf-recent}.

To illustrate once again the accuracy of future measurements at an EIC, 
we also show here a few representative 
projected data points  taken from Fig.~\ref{fig:g1} in the relevant 
$Q^2$ regime around $10\,\mathrm{GeV}^2$ 
for the three different c.m.s.\ energies we consider.
Clearly, measurements of $g_1(x,Q^2)$ at small $x$ will dramatically reduce the uncertainties in the quark
helicity PDFs and, indirectly, through the coupled QCD scale evolution of quarks and gluons 
also in $\Delta g(x,Q^2)$.
At any given $x$, scaling violations for $g_1(x,Q^2)$  
will further constrain $\Delta g(x,Q^2)$. As was already shown in Fig.~\ref{fig:g1},
they are numerically not very pronounced for the optimum DSSV 2014 fit, 
which can be also inferred from Fig.~\ref{fig:g1cloud}, where we show $g_1(x,Q^2)$  
at $Q^2=1$ and $100\,\mathrm{GeV}^2$ in addition to our default scale of $10\,\mathrm{GeV}^2$.
However, each of the alternative fits exhibits a somewhat different $Q^2$ dependence
driven by the uncertainties in the $x$ shapes of the quark and gluon densities.
For $x\gtrsim 0.01$, the scale dependence of $g_1(x,Q^2)$ in the range 
from $Q^2=1$ to $100\,\mathrm{GeV}^2$ is typically larger than the 
uncertainty on $g_1(x,Q^2)$ from present data.

\section{Present status of $\mathbf{\Delta g}$ and Impact of Projected RHIC and EIC data }
%
Before addressing the question of how precisely an EIC will constrain the total gluon 
and quark polarizations in the spin decomposition (\ref{eq:spinsum}), and, 
indirectly, also the total OAM ${\cal{L}}$, it is important to first make a precise assessment of how well 
these quantities are expected to be known by the end of the current experimental programs,
in particular, RHIC spin.
This will set the best possible baseline to judge on the impact a future EIC could have
in the field of QCD spin physics.

%
%
\begin{figure}[tbh!]
\begin{center}
\vspace*{-0.9cm}
\epsfig{figure=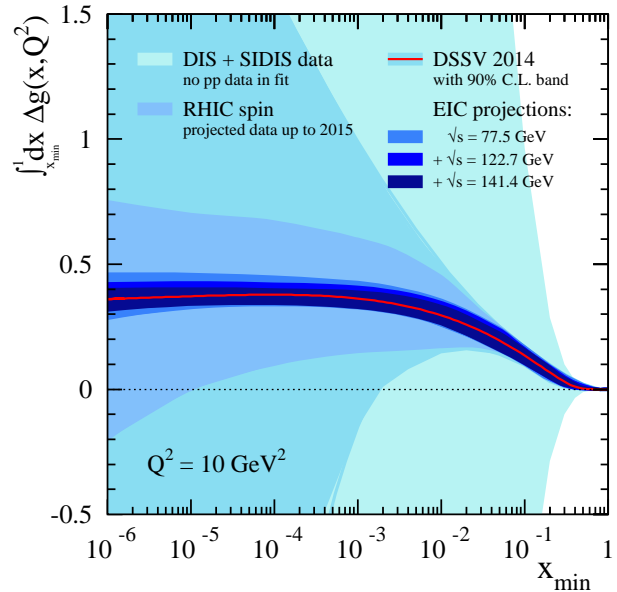,width=0.52\textwidth}
\end{center}
\vspace*{-0.7cm}
\caption{\label{fig:rdg} [color online]
The running integral of the gluon helicity distribution at $Q^2=10\,\mathrm{GeV}^2$
as a function of $x_{\min}$
according to the DSSV 2014 analysis \cite{deFlorian:2014yva} (solid line)
and $90\%$ C.L.\ uncertainty estimates (shaded bands) based on global QCD fits utilizing 
different sets of existing and projected $pp$ and EIC data (see text).}
\end{figure}
Different indicators and measures can be adopted to quantify how well 
the gluon helicity density and the resulting contribution $\Delta g(Q^2)$ to the proton's spin 
are constrained by data.
The standard way to study uncertainties as a function of the parton's momentum fraction $x$ 
at a given $Q^2$ in a global QCD fit to all available data is certainly the most obvious possibility, 
however, it neither provides an immediate idea of the accuracy 
for the phenomenologically interesting $x$-integral (\ref{eq:firstmom})
that is the focus of our study, nor does it
indicate the relevance of the different regions in $x$ probed by the different experiments
used in the fit.

Instead, we choose to present most of our results in terms of
the ``running integral'' of, for instance, the gluon helicity density,
defined analogously to Eq.~(\ref{eq:firstmom}) as
\begin{equation}
\label{eq:running}
\Delta g(Q^2,x_{\min}) \equiv \int_{x_{\min}}^{1} dx\, \Delta g(x,Q^2)\,\,,
\end{equation}
which represents the share of the proton spin (\ref{eq:spinsum}) from gluons 
as a function  of the lower integration limit $x_{\min}$.
Its uncertainty takes into account the non-trivial correlations between the
different regions of $x$ contributing to (\ref{eq:running}).
By varying $x_{\min}$ in (\ref{eq:running}), one can explore 
how low in $x$ -- or, alternatively, how high in $\sqrt{s}$ -- one 
likely needs to go with future experiments
to reduce $x\to 0$ extrapolation uncertainties to a level small enough
to make meaningful statements about how gluons and quarks in the 
proton make up its spin. To study the important question of 
the convergence of (\ref{eq:running})
with $x_{\min}$ in more detail,
we will also compute the contributions to
(\ref{eq:running}) from different bins $[x_{\min},x_{\max}]$ in $x$
in case of $\Delta g$. 

To estimate the impact of past, current, and future data sets on
$\Delta g$ and $\Delta \Sigma$ we proceed in steps. To this end, we
will present uncertainty estimates for various running integrals by
including different data sets one-by-one into our global analysis framework.
As we have mentioned already, to demonstrate the impact an EIC will have on
$\Delta g$ in the future, we should take into account the experimental information
that is expected to become available soon from the RHIC spin program.
Essentially, the RHIC running focusing on longitudinally polarized double-spin
asymmetries has concluded in 2015, and several high impact data
sets are forthcoming, some of which have been presented in preliminary form
at conferences recently \cite{ref:rhic-upcoming}. Since the expected statistical accuracy of these
measurements is already known, we can proceed by generating mock RHIC $pp$ data
with the proper uncertainties to estimate their impact on helicity PDFs,
in particular, the gluon.

Since extending the reach towards smaller values of $x$ is the most important asset to
arrive at a more solid estimate for the integral $\Delta g(Q^2)$, we only focus
on upcoming RHIC measurements which have the best potential to do so, i.e., inclusive jet
and neutral pion data from STAR and PHENIX, respectively, taken at the
highest c.m.s.\ energy of $510\,\mathrm{GeV}$ for $pp$ collisions at RHIC.
In particular, spin asymmetries at forward rapidities will allow one
to probe $x$ values down to about a few times $10^{-3}$, significantly below
the currently covered range in $x$ indicated in Fig.~\ref{fig:deltag}.
More specifically, our projections comprise double spin asymmetries for
\begin{itemize}
\item inclusive jet data at mid rapidity,
\item inclusive $\pi^0$ data at mid rapidity, and
\item inclusive $\pi^0$ data at forward rapidities\,.
\end{itemize}

In Fig.~\ref{fig:rdg} we show the running integral (\ref{eq:running})
of the gluon helicity distribution at $Q^2=10\,\mathrm{GeV}^2$
down to $x_{\min}=10^{-6}$.
Concentrating first on the optimum DSSV 2014 fit (solid line), one
notices that the integral saturates and reaches more than $90\%$ of its
total value $\Delta g(Q^2=10\,\mathrm{GeV}^2)$ already for
$x_{\min}\simeq 10^{-3}$, suggesting 
that most of the gluon spin contribution to the sum (\ref{eq:spinsum})  
stems from momentum fractions above $10^{-3}$.
This is in line with common expectations \cite{Brodsky:1994kg} 
that $\Delta g(x,Q^2)$ behaves like $x g(x,Q^2)$
at sufficiently small values of $x$.
The preferred value for the gluon spin contribution at $Q^2=10\,\mathrm{GeV}^2$ 
in the DSSV 2014 analysis \cite{deFlorian:2014yva} turns out to be rather large, 
about $0.37$, corresponding to roughly $70\%$ of the proton spin.
However, neither the preferred small $x$ behavior, nor the large value for 
$\Delta g(Q^2=10\,\mathrm{GeV}^2)$, are based on actual experimental constraints in the QCD analysis
as is illustrated by the various uncertainty bands shown in Fig.~\ref{fig:rdg}.
Here and in the following, all uncertainty bands correspond to $90\%$ C.L.\ estimates.

The outermost light shaded bands in Fig.~\ref{fig:rdg} represent a global QCD fit based solely
on existing polarized fixed target DIS and SIDIS data \cite{Aidala:2012mv}.
One can safely conclude that these data do not constrain the gluon helicity density
or its running integral by any means.  
The importance of the current RHIC polarized $pp$
data for inclusive jet and $\pi^0$ production \cite{Adare:2008qb,Adare:2008aa,Adamczyk:2012qj,Adamczyk:2013yvv,Adamczyk:2014ozi,Adare:2015oqa} in constraining $\Delta g(x,Q^2)$ down to 
about $x\simeq 0.01$ is exemplified by the second level of shaded bands. 
Still, for smaller values of $x_{min}$ possible variations in the truncated
integral away from its best fit value quickly become very large due to the flexible functional form
adopted in the DSSV framework. At least, the analysis suggests at $90\%$ C.L., that the
truncated integral (\ref{eq:running}) stays positive down to $x_{\min}\simeq 2\times 10^{-3}$,
but nothing can be concluded about the full integral $\Delta g (Q^2)$,
not even its sign.
These uncertainties correspond to those obtained in 
the DSSV 2014 global analysis \cite{deFlorian:2014yva} and
represent our current knowledge. 

Next, we include the projections for the upcoming RHIC $pp$ data as listed above
into the global analysis framework to estimate what can be expected to be known about
$\Delta g (Q^2)$ by the end of RHIC operations and before an EIC will turn on.
Any impact of a future EIC needs to be judged with respect to this baseline.
As can be seen, RHIC can be expected to further constrain the gluon
helicity distribution and, in particular, its running integral, 
mainly by ruling out very extreme, positive or negative, variations at small momentum fractions.
This will be mainly achieved by measurements at forward rapidities.
Even though these data will have no sensitivity below $x\simeq 10^{-3}$,
they considerably limit possible variations within the given very flexible
functional form of the DSSV analysis.
Together with the decreasing weight of the region of very small $x$ 
in the running integral (\ref{eq:running}), the error on the integral down to
$x_{\min}\simeq 10^{-6}$ is significantly reduced but it still covers about twice
the proton spin of $1/2$ and ranges roughly from $-0.25$ to $0.75$.

We are now in a position to focus on the impact of the generated EIC data shown in Fig.~\ref{fig:g1}.
To this end, we successively add the three sets of mock data given in Tab.~\ref{tab:kine}
starting from the lowest c.m.s.\ energy of $\sqrt{s}=77.5\,\mathrm{GeV}$ to the global set of data,
including the RHIC projections just discussed.
The series of the innermost dark shaded bands in Fig.~\ref{fig:rdg} represents the 
corresponding uncertainty estimates.
The results clearly show that for momentum fractions $x \gtrsim 0.05$, DIS, SIDIS,
and RHIC data, as included in the DSSV 2014 analysis \cite{deFlorian:2014yva},
constrain the running integral adequately well whereas an EIC adds very little new here, apart
from a crucial, independent check of what has been learnt so far in that 
particular kinematic region. 

The constraining power of an EIC starts to become increasingly noticeable below 
$x \simeq 0.01$.
Already at around $x_{\min} \simeq 10^{-3}$, EIC DIS data are expected to reduce 
the uncertainties by approximately a factor of 8 as compared to the DSSV 2014 estimate,
and by a factor of 4 relative to the projections based on future RHIC data. 
Below $x_{\min} \simeq 10^{-3}$, EIC data will constrain the flexible functional form
for $\Delta g(x,Q^2)$ adopted both in the DSSV and our analyses in such a way that
uncertainties in the running integral remain constant, excluding extreme variations in the gluon
helicity density at small $x$ still allowed in fits based solely on current or projected RHIC data.

%
\begin{figure}[tbh!]
\begin{center}
\vspace*{-0.9cm}
\epsfig{figure=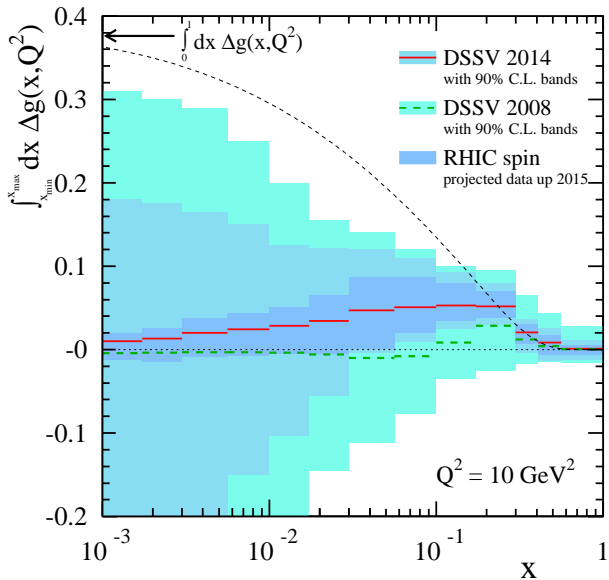,width=0.52\textwidth}
\end{center}
\vspace*{-0.7cm}
\caption{\label{fig:asgard} [color online]
Contributions to $\Delta g(Q^2)$ 
in (\ref{eq:spinsum}) at $Q^2=10\,\mathrm{GeV}^2$ from different $x$ intervals (see text).
The solid and dashed lines represent the results from the DSSV 2014 \cite{deFlorian:2014yva} and
2008 \cite{deFlorian:2008mr,deFlorian:2009vb} best fits, respectively, 
which mainly differ by latest RHIC $pp$ data \cite{Adamczyk:2014ozi,Adare:2015oqa}.
The various shaded bands reflect $90\%$ C.L.\ uncertainty estimates for both fits and the
impact of forthcoming RHIC data. The dotted line indicates the cumulative contribution to
$\Delta g(Q^2)$ for DSSV 2014.}
\end{figure}
Using only EIC data with the lowest c.m.s.\ energy considered in our analysis,
would lead to a determination of the gluon spin contribution to (\ref{eq:spinsum}) to within
a relative uncertainty of about $25\%$. Including data at higher $\sqrt{s}$ would further
reduce relative uncertainties in $\Delta g(Q^2)$ to a level of about $10\%$.
We stress again, that for all our studies, the EIC mock data are generated and randomized around the small 
$x$ extrapolation of the DSSV 2014 best fit (solid line in Fig.~\ref{fig:rdg}).
An EIC would be able to either verify if gluons indeed contribute only very little to 
the running integral below $x_{\min} \simeq 10^{-3}$, as is usually expected \cite{Brodsky:1994kg}, 
or to determine decisively if they exhibit any
different behavior at small $x$. In the latter case, depending on at which $x_{\min}$ the
running integral starts to saturate to its full value, DIS data at the largest possible
c.m.s.\ energy are potentially most relevant to have. 

%
\begin{figure}[tbh!]
\begin{center}
\vspace*{-0.9cm}
\epsfig{figure=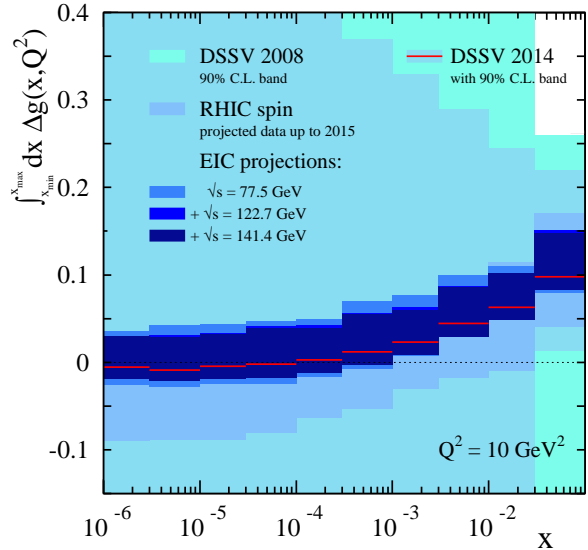,width=0.5\textwidth}
\end{center}
\vspace*{-0.7cm}
\caption{\label{fig:asgard2} [color online]
As in Fig.~\ref{fig:asgard} but now focusing on the small $x$ extrapolation
and the impact of the projected polarized DIS data sets shown in 
Fig.~\ref{fig:g1} on the $90\%$ C.L.\ uncertainty estimates.}
\end{figure}
The relevance that different regions in $x$ have in the running integral shown above
is analyzed more closely in Figs.~\ref{fig:asgard} and \ref{fig:asgard2}.
Figure~\ref{fig:asgard} illustrates the comparative impact of gluons to the spin sum (\ref{eq:spinsum})
in different bins of $x$, four per decade, and the corresponding uncertainties
down to $10^{-3}$ expected to be accessible with forthcoming RHIC $pp$ data.
The solid (dashed) lines in each bin are the estimates computed with the DSSV 2014 
\cite{deFlorian:2014yva} and 2008 \cite{deFlorian:2008mr} fits, respectively. 
The dotted line gives the cumulative contribution to the running integral
(\ref{eq:running}) as was shown in Fig.~\ref{fig:rdg}. The arrow indicates the value of 
full integral $\Delta g(Q^2=10\,\mathrm{GeV}^2$).
In addition, $90\%$ C.L. uncertainty estimates are shown for each individual bin.
We notice that the partial contributions of each bin in $x$ to the running integral (\ref{eq:running})
and the corresponding variations are very strongly correlated to those of the neighboring bins.
As a consequence, the sum of the individual uncertainties from each bin can
exceed the uncertainty of the running integral shown in Fig.~\ref{fig:rdg}. 
Nevertheless, they correctly reflect the relative impact of the different bins in $x$. 
In each bin the uncertainties reflect the synergy of all the different data sets 
included in the global analysis.

The outermost shaded bands in Fig.~\ref{fig:asgard}
correspond to the DSSV 2008 analysis, the middle ones to the
most recent 2014 fit. The optimum gluon helicity distribution in the 2008 fit had an almost vanishing
contribution to the proton spin sum rule \cite{deFlorian:2008mr} and was characterized by a node
in the $x$ region probed by available RHIC data at that time. These features can be easily
gathered also from Fig.~\ref{fig:asgard}. The current 2014 best fit based on the latest
RHIC $pp$ data has a much larger and positive $\Delta g(x,Q^2)$, cf.\ Fig.~\ref{fig:deltag}.
As can be inferred from Fig.~\ref{fig:asgard}, only the $x$ region probed by the $pp$ data
contributes significantly to the integral.
As is expected from Fig.~\ref{fig:rdg}, uncertainties for the old DSSV 2008 analysis are
significantly larger than those for the 2014 fit in all bins of $x$, but both fits cannot constrain 
contributions to the running integral below $x\lesssim 0.01$ in any meaningful way.
Here, uncertainties sharply increase with decreasing $x$.
The innermost, dark shaded uncertainty bands demonstrate the constraining power of 
the projected RHIC data which now, unlike for the previous results, 
exhibit decreasing uncertainties with smaller values of $x$.

In Fig.~\ref{fig:asgard2} we extend the $x$ range down to $10^{-6}$ and include 
bin-by-bin uncertainty estimates based on the three sets of projected DIS data at an EIC.
Compared to Fig.~\ref{fig:asgard}, we now use only 2 bins per decade in $x$.
Below $x \sim 10^{-3}$ the contributions to the integral in each bin 
remain small for the DSSV 2014 best fit (solid line), fluctuating around zero, 
but the corresponding uncertainties continue to increase for both the
2008 and 2014 fits. Uncertainty estimates including also 
the projected RHIC data only slightly increase towards smaller $x$ reflecting the
improvements in the running integral already observed in Fig.~\ref{fig:rdg}.
The constraining power of the EIC DIS data is very significant in the entire $x$ range
and appears to be roughly a constant factor of $2-3$ with respect to the 
bin-by-bin estimates including the projected RHIC data. This is somewhat less
than what was obtained for the running integral at low $x_{\min}$, 
suggesting that scaling violations are more powerful when considered over an
extended range of momentum fractions rather than in a small bin in $x$. In the latter
case much fewer data points are actually providing a strong constraint on allowed
variations in any given bin. Another important factor to consider are the mentioned 
sizable bin-by-bin correlations.  
Similarly, the impact of including more DIS sets at different c.m.s.\ energies is
reduced for the bin-by-bin studies as compared to running integral shown in
Fig.~\ref{fig:rdg}.

\section{Status and Prospects for $\mathbf{\Delta \Sigma}$ and the total OAM contribution}
%
Similarly to what has been discussed in the previous Section in connection
with the gluon helicity density, the running integral of the 
quark singlet $\Delta \Sigma(x,Q^2)$ represents the intrinsic spin contribution of 
all quark flavors in the decomposition of the proton spin (\ref{eq:spinsum}).
Thanks to the direct coupling of the quarks to the probing virtual photon in DIS,
$\Delta \Sigma(x,Q^2)$ is much better constrained by present fixed target data
than the gluon helicity distribution which only enters indirectly through
QCD scale evolution or as an ${\cal{O}}(\alpha_s)$ correction. 
Since $\Delta \Sigma(x,Q^2)$ and $\Delta g(x,Q^2)$ are coupled through the
singlet evolution equations, any constraint from data on either of the two
distributions impacts also the other one.

An extraction of the quark singlet from DIS data on $g_1(x,Q^2)$ also requires to determine
simultaneously two additional flavor non-singlet distributions,
which, if needed, can be all recast into the total contributions from $u$, $d$, and $s$ quarks,
i.e., $\Delta u+\Delta \bar{u}$, $\Delta d + \Delta \bar{d}$, and
$\Delta s + \Delta \bar{s}$ (here we ignore for simplicity any contribution 
from charm and bottom quarks, which play no role for all currently available data).  

%
\begin{figure}[tbh!]
\begin{center}
\vspace*{-0.9cm}
\epsfig{figure=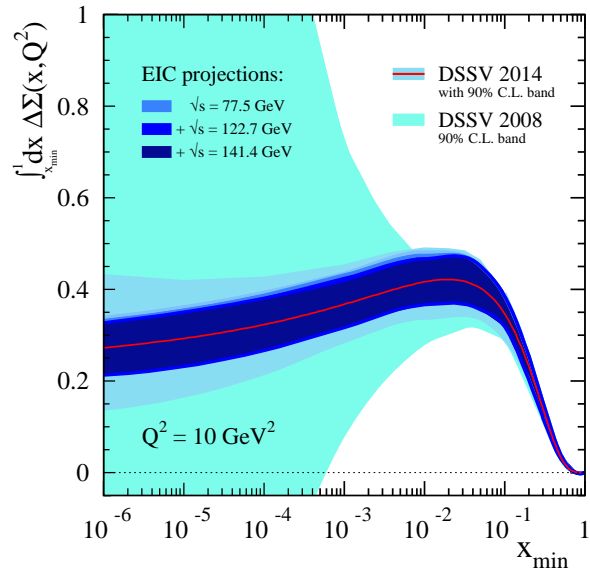,width=0.5\textwidth}
\end{center}
\vspace*{-0.7cm}
\caption{\label{fig:rds} [color online]
Similar to Fig.~\ref{fig:rdg} but now for the running integral of the quark singlet 
helicity density $\Delta \Sigma$. The solid line corresponds to the optimum fit of
DSSV 2014 \cite{deFlorian:2014yva}. $90\%$ C.L.\ uncertainty estimates (shaded bands) are shown
for the DSSV 2008 and 2014 analyses and after including the different sets of projected EIC data.}
\end{figure}
The $x$-integrals of the two non-singlet combinations are usually assumed to be
related to the hyperon decay constants $F$ and $D$ within some uncertainties,
which provide some indirect constraint for the currently unmeasured small $x$ region
below a few times $10^{-3}$.
Since we only wish to focus on the quark flavor singlet in this paper, we adopt these
constraints in the same way as was done in the DSSV global analyses. A more thorough
analysis of flavor separated quark helicity densities and their uncertainties
will be conducted once a fully updated suite of reliable sets of fragmentation functions
becomes available to revisit our previous impact study of SIDIS data in Ref.~\cite{Aschenauer:2012ve}.
Then we will also explore how well
an EIC can challenge the constraints imposed on the quark sector by the hyperon decays, 
one of which being related to the Bjorken sum rule, the other, more important one mainly to
the amount of strangeness polarization in the nucleon.

In Fig.~\ref{fig:rds} we show $90\%$ C.L. estimates for the running integral
of $\Delta \Sigma(x,Q^2)$ as a function of $x_{\min}$ for $Q^2=10\,\mathrm{GeV}^2$
for fits including different sets of existing and projected data.
One should notice that the vertical axis here covers only half of the range
shown for the running integral of $\Delta g$ in Fig.~\ref{fig:rdg}. Also,
$\Delta \Sigma(Q^2)$ enters the spin sum rule (\ref{eq:spinsum}) with a
factor of $1/2$ relative to the gluon spin contribution.

The outermost shaded band represents uncertainties as present in the original 
DSSV global analysis from 2008 \cite{deFlorian:2008mr}. They appear to be very significant
for $x_{min}\lesssim 10^{-3}$. 
As usual, the solid line shows the optimum fit of DSSV 2014 extrapolated down in $x$.
The corresponding uncertainties are much reduced as compared to the 2008 analysis
due to including additional DIS data from the COMPASS collaboration 
and, indirectly, through constraints on $\Delta g$ from RHIC $pp$ 
data \cite{Adamczyk:2014ozi,Adare:2015oqa}. 

%
\begin{figure}[thb!]
\begin{center}
\vspace*{-0.9cm}
\epsfig{figure=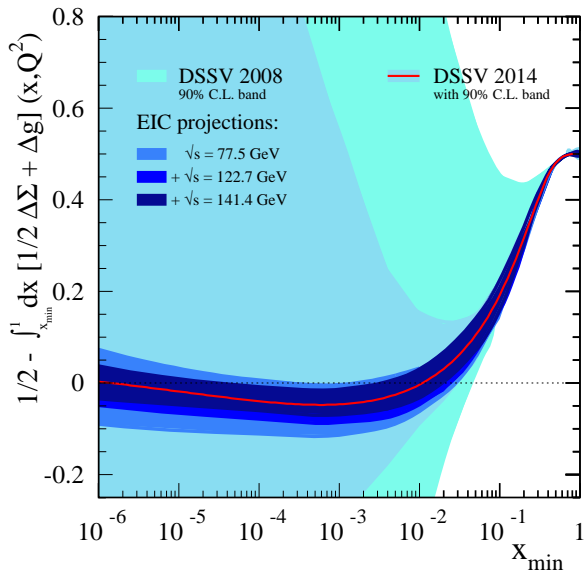,width=0.5\textwidth}
\end{center}
\vspace*{-0.7cm}
\caption{\label{fig:rdl} [color online]
As in Fig.~\ref{fig:rds} but now showing the contribution from the combined quark and
gluon OAM ${\cal{L}}(Q^2)$ using Eq.~(\ref{eq:spinsum}) and the results obtained
for the running integrals of $\Delta g$ and $\Delta \Sigma$. As before, 
$90\%$ C.L.\ uncertainty estimates (shaded bands) illustrate the impact of different projected EIC data sets.}
\end{figure}
The set of three innermost shaded bands illustrates the significant impact of an 
EIC from a series of global fits that successively include the projected DIS data sets starting from
the one corresponding to the lowest c.m.s.\ energy.  The addition of the sets with increasing 
c.m.s.\ energy has less impact on the uncertainties than for the gluon helicity distribution
shown in Fig.~\ref{fig:rdg}. 
For $x_{\min}=10^{-6}$ and with an EIC, one expects from our studies 
to control the value of $\Delta \Sigma$ to within about $15\%$ which is somewhat 
worse than what can be achieved for $\Delta g$.
Also, the convergence of the running integral for $\Delta \Sigma$ for the
DSSV 2014 best fit is much slower than for the gluon density shown in Fig.~\ref{fig:rdg}.
This can be mainly attributed to the small $x$ behavior of the strangeness helicity
distribution and the above mentioned imposed constraints from the hyperon decay constants $F$
and $D$. An EIC will be able to verify the validity of these assumptions by determining
the sea quark distributions, in particular, $\Delta s(x,Q^2)$, 
down to unprecedentedly small values of $x$ 
from a series of SIDIS measurements with identified pions and kaons.
Once this is achieved, we expect that a future combined global analysis of SIDIS and DIS
data will also lead to a much improved uncertainty estimate on the integral of $\Delta \Sigma$ 
compared to what is presented here. 
 
As we have demonstrated in Figs.~\ref{fig:rdg} and \ref{fig:rds}, 
an EIC will deliver a precise picture of the intrinsic
spin contributions of quarks and gluons to within at least $10-15\%$ relative
uncertainties. This information can be used along with the proton's spin
decomposition (\ref{eq:spinsum}) to estimate how much is left for the combined
contribution ${\cal{L}}(Q^2)$ from the orbital motion of quarks and gluons.
This is illustrated in terms of the running integral of ${\cal{L}}(Q^2)$ 
in Fig.~\ref{fig:rdl} along with our usual set of
uncertainty estimates based on the different sets of existing and projected experimental data.

The solid line gives the estimate for ${\cal{L}}(Q^2)$ for $Q^2=10\,\mathrm{GeV}^2$
based on extrapolating the best fit results of DSSV 2014 for $\Delta g(x,Q^2)$ and $\Delta\Sigma(x,Q^2)$ down to $x_{\min}$ and subtracting them off the total proton spin of $1/2$.
As it turns at, at this particular value of $Q^2$, the DSSV 2014 result converges to
basically a zero net contribution from OAM to the spin sum rule (\ref{eq:spinsum})
but within huge uncertainties. The latter is given for both the DSSV 2008 and 2014
global fits by the outermost and middle shaded bands, respectively.
As one can already anticipate from our studies performed in Figs.~\ref{fig:rdg} and \ref{fig:rds}, 
an EIC will yield an excellent indirect constraint on ${\cal{L}}(Q^2)$ by combining the
then available precise information on $\Delta g$ and $\Delta \Sigma$.
As can be inferred from the corresponding uncertainty estimates (set of innermost
shaded bands), one can expect to constrain the net OAM contribution of
quarks and gluons to within about $\pm 0.05$ when all three sets of projected 
DIS data are combined. Of course, only a set of dedicated measurements 
at an EIC can reveal a detailed, hopefully flavor separated,
quantitative picture of the orbital motion of quarks and gluons.
For instance, even if the net OAM contribution turns out to be small as
for the DSSV 2014 best fit, there might be significant cancellations among
the individual quark flavors and the gluon. In addition, $\Delta g(Q^2)$ in
(\ref{eq:spinsum}) evolves logarithmically with $Q^2$ such that any increase
has to be compensated by a corresponding decrease of ${\cal{L}}(Q^2)$, and vice versa,
to always arrive at $1/2$.
We recall that $\Delta \Sigma(Q^2)$ is a renormalization group invariant at LO and,
in the $\overline{\mathrm{MS}}$ scheme, evolves only very slowly with $Q^2$.
However, separating quark and gluon OAM in Eq.~(\ref{eq:spinsum}) experimentally involves
determinations of twist-three generalized parton as well as quantum 
phase-space Wigner distributions which will be very challenging and still needs
further theoretical work and perhaps lattice QCD studies, see, for instance,
Refs.~\cite{Leader:2013jra} and \cite{ref:oam}.

\section{Summary and Conclusions}
%
The Electron Ion Collider project constitutes a versatile and vast program to considerably
deepen our knowledge of the inner workings of nucleons and nuclei and the underlying dynamics
and interactions of quarks and gluons as governed by Quantum Chromodynamics.
We have presented a detailed account of what can be achieved in the field of Spin Physics,
in particular, the complex interplay of quark and gluon spins and their orbital motions
yielding the known spin $1/2$ of the proton, for which we still lack a detailed quantitative
understanding despite more than thirty years of intense research.

In the present study, we have considerably updated and refined our
previous analysis on the impact of future EIC polarized DIS data 
in connection to our understanding of the gluon spin and its share in the proton spin budget. 
Throughout, we have explored the consequences of the much larger degree of gluon polarization in 
the nucleon than it has hitherto been acknowledged, as implied by the most recent RHIC results.
In addition, we have used realistic projections for several forthcoming sets of 
RHIC data to estimate to what extent they further constrain the gluon helicity distribution
by the end of the RHIC program. This baseline was used to quantify the expected impact
of a future polarized DIS program at an EIC.

To this end, we have focused on providing accurate estimates of uncertainties for
the running integrals as a function of the minimum momentum fraction
for the gluon and quark singlet helicity distributions.
Results were obtained in the context of global QCD analyses based on
current and various projected sets of data and using the
robust Lagrange multiplier technique for all error estimations.
In case of the gluon helicity distribution we also provided detailed
uncertainty studies bin-by-bin in momentum fraction to establish where
the running integral receives major contributions and to determine 
below which value of $x$ one can expect the integral to converge.

We find that the RHIC spin program still has quite some potential in reducing
current uncertainties in the gluon helicity distribution by ruling out extreme
variations in its shape. Nevertheless, one cannot expect to arrive at any 
meaningful and reliable estimate of the gluon contribution in the proton spin decomposition due
to remaining substantial extrapolation uncertainties from the unmeasured region
of small momentum fractions. 
Even under very conservative assumptions for its performance, an EIC
with a sufficiently high enough center-of-mass system energy to reach deep into
the so far unexplored kinematic regime,
will decisively constrain both the gluon and the singlet quark helicity
densities and, at last, elevate the field of QCD spin physics into an precision era.
For the corresponding $x$-integrals relevant for the proton's spin decomposition
we expect to achieve relative uncertainties of about $10\%$. For the first time,
this will also lead to a precise indirect estimate of the remaining contribution
to the proton spin from the combined orbital motion of quarks and gluons, well before
an EIC will be able to determine them directly.
Estimates obtained by extrapolating our current best knowledge of helicity densities, without EIC data,
to small momentum fractions point to a surprisingly small net orbital angular momentum 
in the proton's spin budget at intermediate scales $Q^2$, albeit within very large uncertainties.

\section*{Acknowledgments}
%
We are grateful to A.\ Bazilevsky and C.\ Gagliardi for providing us with the
projections for the RHIC spin program.
R.S.\ and M.S.\ are grateful to Brookhaven National Lab for hospitality while
completing this study. 
This work was partially supported by CONICET, ANPCyT, UBACyT, the
U.S.\ Department of Energy under Contract No.~{DE-SC0012704}, and
the Institutional Strategy of the University of T\"{u}bingen (DFG, ZUK 63).


\end{document}